\documentclass[twocolumn,letterpaper,aps,prl,longbibliography,superscriptaddress,showpacs,floatfix]{revtex4-1}

\usepackage{graphicx}	
\usepackage{xspace}	

\newcommand{\pt}{\mbox{$p_T$}\xspace}

\newcommand{\sqs}{\mbox{$\sqrt{s}$}\xspace}

\newcommand{\pp}{\mbox{$p$$+$$p$}\xspace}

\newcommand{\piz}{\mbox{$\pi^{0}$}\xspace}
\newcommand{\ptp}{\mbox{$p^\uparrow$$+$$p$}\xspace}

\newcommand{\gevc}{~{\rm GeV}/c}

\newcommand{\gev}{~{\rm GeV}}
\newcommand{\mevcc}{~{\rm MeV}/c^2}

\def\Fig#1{Figure~\ref{#1}}
\def\fig#1{Fig.~\ref{#1}}
\def\eq#1{Eq.~(\ref{#1})}

\begin{document}



\title{Probing gluon spin-momentum correlations in transversely 
polarized protons through midrapidity isolated direct photons in 
$p^\uparrow+p$ collisions at $\sqrt{s}=200$ GeV}


\newcommand{\abilene}{Abilene Christian University, Abilene, Texas 79699, USA}
\newcommand{\augie}{Department of Physics, Augustana University, Sioux Falls, South Dakota 57197, USA}
\newcommand{\banaras}{Department of Physics, Banaras Hindu University, Varanasi 221005, India}
\newcommand{\barc}{Bhabha Atomic Research Centre, Bombay 400 085, India}
\newcommand{\baruch}{Baruch College, City University of New York, New York, New York, 10010 USA}
\newcommand{\bnlcoll}{Collider-Accelerator Department, Brookhaven National Laboratory, Upton, New York 11973-5000, USA}
\newcommand{\bnlphys}{Physics Department, Brookhaven National Laboratory, Upton, New York 11973-5000, USA}
\newcommand{\caucr}{University of California-Riverside, Riverside, California 92521, USA}
\newcommand{\charlesczech}{Charles University, Ovocn\'{y} trh 5, Praha 1, 116 36, Prague, Czech Republic}
\newcommand{\cns}{Center for Nuclear Study, Graduate School of Science, University of Tokyo, 7-3-1 Hongo, Bunkyo, Tokyo 113-0033, Japan}
\newcommand{\colorado}{University of Colorado, Boulder, Colorado 80309, USA}
\newcommand{\columbia}{Columbia University, New York, New York 10027 and Nevis Laboratories, Irvington, New York 10533, USA}
\newcommand{\czechtech}{Czech Technical University, Zikova 4, 166 36 Prague 6, Czech Republic}
\newcommand{\debrecen}{Debrecen University, H-4010 Debrecen, Egyetem t{\'e}r 1, Hungary}
\newcommand{\elte}{ELTE, E{\"o}tv{\"o}s Lor{\'a}nd University, H-1117 Budapest, P{\'a}zm{\'a}ny P.~s.~1/A, Hungary}
\newcommand{\eszterhazy}{Eszterh\'azy K\'aroly University, K\'aroly R\'obert Campus, H-3200 Gy\"ongy\"os, M\'atrai \'ut 36, Hungary}
\newcommand{\ewha}{Ewha Womans University, Seoul 120-750, Korea}
\newcommand{\famu}{Florida A\&M University, Tallahassee, FL 32307, USA}
\newcommand{\fsu}{Florida State University, Tallahassee, Florida 32306, USA}
\newcommand{\gsu}{Georgia State University, Atlanta, Georgia 30303, USA}
\newcommand{\hiroshima}{Hiroshima University, Kagamiyama, Higashi-Hiroshima 739-8526, Japan}
\newcommand{\howard}{Department of Physics and Astronomy, Howard University, Washington, DC 20059, USA}
\newcommand{\ihepprot}{IHEP Protvino, State Research Center of Russian Federation, Institute for High Energy Physics, Protvino, 142281, Russia}
\newcommand{\illuiuc}{University of Illinois at Urbana-Champaign, Urbana, Illinois 61801, USA}
\newcommand{\inrras}{Institute for Nuclear Research of the Russian Academy of Sciences, prospekt 60-letiya Oktyabrya 7a, Moscow 117312, Russia}
\newcommand{\instpasczech}{Institute of Physics, Academy of Sciences of the Czech Republic, Na Slovance 2, 182 21 Prague 8, Czech Republic}
\newcommand{\isu}{Iowa State University, Ames, Iowa 50011, USA}
\newcommand{\jaea}{Advanced Science Research Center, Japan Atomic Energy Agency, 2-4 Shirakata Shirane, Tokai-mura, Naka-gun, Ibaraki-ken 319-1195, Japan}
\newcommand{\jeonbuk}{Jeonbuk National University, Jeonju, 54896, Korea}
\newcommand{\kek}{KEK, High Energy Accelerator Research Organization, Tsukuba, Ibaraki 305-0801, Japan}
\newcommand{\korea}{Korea University, Seoul 02841, Korea}
\newcommand{\kurchatov}{National Research Center ``Kurchatov Institute", Moscow, 123098 Russia}
\newcommand{\kyoto}{Kyoto University, Kyoto 606-8502, Japan}
\newcommand{\lawllnl}{Lawrence Livermore National Laboratory, Livermore, California 94550, USA}
\newcommand{\losalamos}{Los Alamos National Laboratory, Los Alamos, New Mexico 87545, USA}
\newcommand{\lund}{Department of Physics, Lund University, Box 118, SE-221 00 Lund, Sweden}
\newcommand{\lyon}{IPNL, CNRS/IN2P3, Univ Lyon, Université Lyon 1, F-69622, Villeurbanne, France}
\newcommand{\maryland}{University of Maryland, College Park, Maryland 20742, USA}
\newcommand{\mass}{Department of Physics, University of Massachusetts, Amherst, Massachusetts 01003-9337, USA}
\newcommand{\michigan}{Department of Physics, University of Michigan, Ann Arbor, Michigan 48109-1040, USA}
\newcommand{\muhlenberg}{Muhlenberg College, Allentown, Pennsylvania 18104-5586, USA}
\newcommand{\nara}{Nara Women's University, Kita-uoya Nishi-machi Nara 630-8506, Japan}
\newcommand{\natmephi}{National Research Nuclear University, MEPhI, Moscow Engineering Physics Institute, Moscow, 115409, Russia}
\newcommand{\newmex}{University of New Mexico, Albuquerque, New Mexico 87131, USA}
\newcommand{\nmsu}{New Mexico State University, Las Cruces, New Mexico 88003, USA}
\newcommand{\northcg}{Physics and Astronomy Department, University of North Carolina at Greensboro, Greensboro, North Carolina 27412, USA}
\newcommand{\ohio}{Department of Physics and Astronomy, Ohio University, Athens, Ohio 45701, USA}
\newcommand{\ornl}{Oak Ridge National Laboratory, Oak Ridge, Tennessee 37831, USA}
\newcommand{\orsay}{IPN-Orsay, Univ.~Paris-Sud, CNRS/IN2P3, Universit\'e Paris-Saclay, BP1, F-91406, Orsay, France}
\newcommand{\peking}{Peking University, Beijing 100871, People's Republic of China}
\newcommand{\pnpi}{PNPI, Petersburg Nuclear Physics Institute, Gatchina, Leningrad region, 188300, Russia}
\newcommand{\pusan}{Pusan National University, Pusan 46241, Korea}
\newcommand{\riken}{RIKEN Nishina Center for Accelerator-Based Science, Wako, Saitama 351-0198, Japan}
\newcommand{\rikjrbrc}{RIKEN BNL Research Center, Brookhaven National Laboratory, Upton, New York 11973-5000, USA}
\newcommand{\rikkyo}{Physics Department, Rikkyo University, 3-34-1 Nishi-Ikebukuro, Toshima, Tokyo 171-8501, Japan}
\newcommand{\saispbstu}{Saint Petersburg State Polytechnic University, St.~Petersburg, 195251 Russia}
\newcommand{\seoulnat}{Department of Physics and Astronomy, Seoul National University, Seoul 151-742, Korea}
\newcommand{\stonybrkc}{Chemistry Department, Stony Brook University, SUNY, Stony Brook, New York 11794-3400, USA}
\newcommand{\stonycrkp}{Department of Physics and Astronomy, Stony Brook University, SUNY, Stony Brook, New York 11794-3800, USA}
\newcommand{\tenn}{University of Tennessee, Knoxville, Tennessee 37996, USA}
\newcommand{\texsu}{Texas Southern University, Houston, TX 77004, USA}
\newcommand{\titech}{Department of Physics, Tokyo Institute of Technology, Oh-okayama, Meguro, Tokyo 152-8551, Japan}
\newcommand{\tsukuba}{Tomonaga Center for the History of the Universe, University of Tsukuba, Tsukuba, Ibaraki 305, Japan}
\newcommand{\vandy}{Vanderbilt University, Nashville, Tennessee 37235, USA}
\newcommand{\weizmann}{Weizmann Institute, Rehovot 76100, Israel}
\newcommand{\wigner}{Institute for Particle and Nuclear Physics, Wigner Research Centre for Physics, Hungarian Academy of Sciences (Wigner RCP, RMKI) H-1525 Budapest 114, POBox 49, Budapest, Hungary}
\newcommand{\yonsei}{Yonsei University, IPAP, Seoul 120-749, Korea}
\newcommand{\zagreb}{Department of Physics, Faculty of Science, University of Zagreb, Bijeni\v{c}ka c.~32 HR-10002 Zagreb, Croatia}
\affiliation{\abilene}
\affiliation{\augie}
\affiliation{\banaras}
\affiliation{\barc}
\affiliation{\baruch}
\affiliation{\bnlcoll}
\affiliation{\bnlphys}
\affiliation{\caucr}
\affiliation{\charlesczech}
\affiliation{\cns}
\affiliation{\colorado}
\affiliation{\columbia}
\affiliation{\czechtech}
\affiliation{\debrecen}
\affiliation{\elte}
\affiliation{\eszterhazy}
\affiliation{\ewha}
\affiliation{\famu}
\affiliation{\fsu}
\affiliation{\gsu}
\affiliation{\hiroshima}
\affiliation{\howard}
\affiliation{\ihepprot}
\affiliation{\illuiuc}
\affiliation{\inrras}
\affiliation{\instpasczech}
\affiliation{\isu}
\affiliation{\jaea}
\affiliation{\jeonbuk}
\affiliation{\kek}
\affiliation{\korea}
\affiliation{\kurchatov}
\affiliation{\kyoto}
\affiliation{\lawllnl}
\affiliation{\losalamos}
\affiliation{\lund}
\affiliation{\lyon}
\affiliation{\maryland}
\affiliation{\mass}
\affiliation{\michigan}
\affiliation{\muhlenberg}
\affiliation{\nara}
\affiliation{\natmephi}
\affiliation{\newmex}
\affiliation{\nmsu}
\affiliation{\northcg}
\affiliation{\ohio}
\affiliation{\ornl}
\affiliation{\orsay}
\affiliation{\peking}
\affiliation{\pnpi}
\affiliation{\pusan}
\affiliation{\riken}
\affiliation{\rikjrbrc}
\affiliation{\rikkyo}
\affiliation{\saispbstu}
\affiliation{\seoulnat}
\affiliation{\stonybrkc}
\affiliation{\stonycrkp}
\affiliation{\tenn}
\affiliation{\texsu}
\affiliation{\titech}
\affiliation{\tsukuba}
\affiliation{\vandy}
\affiliation{\weizmann}
\affiliation{\wigner}
\affiliation{\yonsei}
\affiliation{\zagreb}
\author{U.A.~Acharya} \affiliation{\gsu} 
\author{C.~Aidala} \affiliation{\michigan} 
\author{Y.~Akiba} \email[PHENIX Spokesperson: ]{akiba@rcf.rhic.bnl.gov} \affiliation{\riken} \affiliation{\rikjrbrc} 
\author{M.~Alfred} \affiliation{\howard} 
\author{V.~Andrieux} \affiliation{\michigan} 
\author{N.~Apadula} \affiliation{\isu} 
\author{H.~Asano} \affiliation{\kyoto} \affiliation{\riken} 
\author{B.~Azmoun} \affiliation{\bnlphys} 
\author{V.~Babintsev} \affiliation{\ihepprot} 
\author{N.S.~Bandara} \affiliation{\mass} 
% ...
\collaboration{PHENIX Collaboration} \noaffiliation


\date{\today}

\begin{abstract}


Studying spin-momentum correlations in hadronic collisions offers a 
glimpse into a three-dimensional picture of proton structure.  The 
transverse single-spin asymmetry for midrapidity isolated direct photons 
in $p^\uparrow+p$ collisions at $\sqrt{s}=200$ GeV is measured with 
the PHENIX detector at the Relativistic Heavy Ion Collider (RHIC). 
Because direct photons in particular are produced from the hard 
scattering and do not interact via the strong force, this measurement is 
a clean probe of initial-state spin-momentum correlations inside the 
proton and is in particular sensitive to gluon interference effects 
within the proton.  This is the first time direct photons have been used 
as a probe of spin-momentum correlations at RHIC.  The uncertainties on 
the results are a fifty-fold improvement with respect to those of the 
one prior measurement for the same observable, from the Fermilab E704 
experiment.  These results constrain gluon spin-momentum correlations in 
transversely polarized protons.

\end{abstract}

\maketitle



Unlike lepton-hadron scattering, proton-proton collisions are sensitive 
to gluon scattering at leading order.  Direct photons are produced 
\textit{directly} in the hard scattering of partons and, because they do 
not interact via the strong force, are a phenomenologically clean probe 
of the structure of the proton. At large transverse momentum, direct 
photons are produced at leading order via the quantum chromodynamics 
(QCD) 2-to-2 hard scattering subprocesses quark-gluon Compton scattering 
(\( g + q \rightarrow \gamma + q \)) and quark-antiquark annihilation 
(\( \bar{q} + q \rightarrow \gamma + g\)).  Compton scattering dominates 
at midrapidity~\cite{Adare:2010yw} because the proton is being probed at 
moderate longitudinal momentum fraction, $x$, where gluons are the 
primary constituents of the proton.  Thus midrapidity direct photon 
measurements are a clean probe of gluon structure within the proton.

Transverse single-spin asymmetries (TSSAs) in hadronic collisions are 
sensitive to various spin-momentum correlations, i.e.~correlations between 
the directions of the spin and momentum of partons and/or hadrons involved 
in a scattering event.  In collisions between one transversely polarized 
proton and one unpolarized proton, the TSSA describes the azimuthal-angular 
dependence of particle production relative to the transverse polarization 
direction.  TSSAs have been measured to be as large as 40\% in forward 
charged pion production~\cite{Klem:1976ui, Adams:1991cs, Allgower:2002qi, 
Arsene:2008aa} and significantly nonzero forward neutral pion asymmetries 
have been measured with transverse momentum up to 
\( p_T \approx 7\gevc \)~\cite{Adam:2020jjg}.  In this context, $p_T$ serves 
as proxy for a hard-scattering energy ($Q$) that is well into the 
perturbative regime of QCD.  Next-to-leading-order perturbative QCD 
calculations, which only include effects from high energy parton scattering 
predict that these asymmetries should be small and fall off as 
$m_q/Q$~\cite{Kane:1978nd}, where $m_q$ is the bare mass of the quark.  
Thus, to explain these large TSSAs, they must be considered in the context 
of the dynamics present in proton-proton collisions that cannot be 
calculated perturbatively, namely dynamics describing proton structure 
and/or the process of hadronization.

One approach toward explaining the large measured TSSAs is through 
transverse-momentum-dependent (TMD) functions.  These functions depend 
on the soft-scale-parton transverse momentum, $k_T$, in addition to 
the partonic longitudinal momentum fraction $x$ and $Q$, where 
$k_T \ll Q$.  TMD functions can be directly extracted from measurements that 
are sensitive to two momentum scales, such as semi-inclusive deep-inelastic 
scattering (SIDIS) where the angle and energy of the scattered electron can 
be used to directly measure the hard-scale $Q$ and the transverse 
momentum of the measured hadron relates to the soft scales $k_T$ of TMD 
parton distribution functions (PDFs) and fragmentation functions.  The 
Sivers function is a PDF that describes the structure of the transversely 
polarized proton and correlates the transverse spin of the proton and 
$k_T$ of the parton within it~\cite{Sivers:1989cc}. The quark Sivers function 
has been extracted through polarized SIDIS measurements, but the gluon 
Sivers function has remained comparatively less constrained because SIDIS is 
not sensitive to gluons at leading order~\cite{Adolph:2017pgv}.  The direct 
photon TSSA in proton-proton collisions has been shown to be sensitive to 
the gluon Sivers function~\cite{Godbole:2018mmh}, but the $ k_T $ moment of 
TMD functions must be used to apply these functions to the single-scale 
inclusive TSSAs measured in proton-proton collisions.

Twist-3 correlation functions are another approach toward describing 
TSSAs.  Unlike TMD functions, collinear twist-3 correlation functions 
depend only on a single scale, the hard scale $Q$.  Twist-3 functions 
describe spin-momentum correlations generated by the quantum mechanical 
interference between scattering off of one parton versus scattering off 
of two.  There are two different types: the quark-gluon-quark 
(\textit{qgq}) correlation functions and the trigluon (\textit{ggg}) 
correlation function.  In the context of proton structure, \textit{qgq} 
correlation functions describe the interference between scattering off 
of a single quark in the proton versus scattering off of a quark, which 
carries the same flavor and the same momentum fraction and an additional 
gluon.  Analogously, the trigluon correlation describes the interference 
between scattering off of one gluon in the proton versus scattering off 
of two.  Additional twist-3 collinear correlation functions describing 
spin-momentum correlations in the process of hadronization also exist, 
but are not relevant to the production of direct photons. Collinear 
twist-3 functions have been shown to be related to the $k_T$ moment of 
TMD functions~\cite{Boer:2003cm, Ji:2006ub}.  For example, the 
Efremov-Teryaev-Qiu-Sterman (ETQS) function is a \textit{qgq} 
correlation function for the polarized 
proton~\cite{Efremov:1984ip,Qiu:1991wg, Qiu:1998ia} that is related to 
the $k_T$ moment of the Sivers TMD PDF. The ETQS function has also been 
extracted from fits to inclusive TSSAs in proton-proton 
collisions~\cite{Kanazawa:2014dca,Cammarota:2020qcw}, and the forward 
direct photon TSSA has been suggested to be dominated by this ETQS 
function~\cite{Kanazawa:2014nea}. The fact that both TMD and collinear 
twist-3 functions are nonzero reflects that scattering partons do in 
fact interact with the color fields present inside the proton, which 
goes beyond traditional assumptions present in hadronic collision 
studies.

Multiple observables can provide sensitivity to the \textit{ggg} 
correlation function.  Midrapidity inclusive hadron TSSA measurements 
are sensitive to gluon spin-momentum correlations in the proton but also 
include potential effects from hadronization and final-state color 
interactions.  Heavy flavor production at the Relativistic Heavy Ion 
Collider (RHIC) is dominated by gluon-gluon fusion and thus particularly 
sensitive to gluons in the proton. A heavy flavor hadron TSSA 
measurement~\cite{Aidala:2017pum} has been used to estimate the trigluon 
correlation function in the transversely polarized proton assuming no 
effects from hadronization or final-state color 
interactions~\cite{Koike:2011mb}.  The midrapidity isolated direct 
photon TSSA is instead a clean probe of the trigluon correlation 
function because it is insensitive to hadronization effects as well as 
final-state color interactions~\cite{Koike:2011nx}.

The only previously published direct photon TSSA measurement is the 
Fermilab E704 result, which used a $200\gevc$ polarized proton beam on an 
unpolarized proton target ($\sqrt{s}=19.4$~\gev).  It was found to be 
consistent with zero to within 20\% for $2.5 < p_T^\gamma < 3.1 
\gevc$~\cite{Adams:1995gg}. The PHENIX results presented in this Letter 
measure photons with $p_T^\gamma>5\gevc$ with total uncertainties up to 
a factor of 50 smaller than the E704 measurements.  This measurement 
will constrain trigluon correlations in transversely polarized protons.

The presented direct photon measurement was performed with the PHENIX 
experiment in the central rapidity region $|\eta|<0.35$, using \ptp 
collisions at \sqs=200\gev.  The data set was collected in 2015 and 
corresponds to an integrated luminosity of approximately 60 pb$^{-1}$.  
Direct photons were reconstructed using similar techniques to a 
previously published direct photon cross section result at \sqs = 
200\gev~\cite{Adare:2012yt}. The asymmetry was measured with 
transversely polarized proton beams at RHIC where the clockwise and 
counter-clockwise beams had an average polarization of $0.58\pm 0.02$ 
and $0.60 \pm 0.02$, respectively~\cite{polarimetry}. Collisions between 
bunches are spaced 106 ns apart and the polarization direction changes 
bunch-to-bunch such that two statistically independent asymmetries can 
be measured with the same particle yields through sorting them by the 
polarization direction in one beam, effectively averaging over the 
polarization in the other beam.  These two independent measurements 
serve as a cross check and are averaged together to calculate the final 
asymmetry.

The PHENIX central detector comprises two nearly-back-to-back arms 
each covering $\Delta\phi=\pi/2$ in azimuth and $|\eta|<0.35$ in 
pseudorapidity.  Photons are identified through clusters in the 
electromagnetic calorimeter (EMCal), which has two detector arms: the 
west and the east.  The west arm comprises four sectors of sampling 
lead-scintillator (PbSc) calorimeters with granularity 
$\delta\phi \times \delta \eta =0.011 \times 0.011$ and the east arm 
comprises two more PbSc sectors along with two sectors of \v{C}erenkov 
lead-glass (PbGl) calorimeters with granularity 
$\delta\phi \times \delta \eta =0.008 \times 0.008$~\cite{Aphecetche:2003zr}. 

The PHENIX central tracking system uses pad chambers and a drift chamber to 
measure the position of charged particle tracks~\cite{Adcox:2003zp}.  The 
beam-beam counters (BBC) are far-forward arrays of quartz \v{C}erenkov 
radiators that cover the full azimuth and 
\( 3.0 < |\eta| < 3.9 \)~\cite{InnerPHENIX}. They measure the position of 
the vertex in the beam direction, for which a 30 cm vertex cut around the 
nominal collision point is applied. The minimum-bias trigger fires on 
crossings where at least one charged particle is measured in each arm of the 
BBC. Events with high-\pt photons are selected through an EMCal-based 
high-energy photon trigger that is taken in coincidence with this 
minimum-bias trigger.

All photons used in the asymmetry calculation are required to pass the 
following cuts.  A shower shape cut selects clusters whose energy 
distribution is consistent with a parameterized profile from a photon 
shower. This reduces the contribution of clusters from hadrons along 
with merged photons from high energy \piz decays, which resolve as a 
single cluster in the EMCal.  A time-of-flight cut suppresses the 
contribution of EMCal noise, where the timing of the cluster is measured 
by the EMCal and the time zero reference of the event is provided by the 
BBC.  A charged-track-veto cut eliminates clusters that geometrically 
match with a charged track and uses the track position measured directly 
in front of the EMCal.  This cut reduces the background from electrons 
as well as charged hadrons that were not eliminated by the shower shape 
cut.

Direct photon candidates are also required to pass tagging cuts that 
reduce the hadronic decay background by eliminating photons that are 
tagged as coming from either \( \piz \rightarrow \gamma \gamma \) or 
\( \eta \rightarrow \gamma \gamma \) decays. The candidate direct photon is 
matched with a partner photon in the same event and same EMCal arm, which 
has passed a minimum-energy cut of 0.5\gev.  A photon is considered 
tagged as coming from a \( \piz \rightarrow \gamma \gamma \) (\( \eta 
\rightarrow \gamma \gamma \)) decay if it is matched into a photon pair 
with invariant mass \( 105 < M_{\gamma\gamma} < 165 \mevcc\) (\( 480 < 
M_{\gamma\gamma} < 620 \mevcc \)), which corresponds roughly to a $\pm 2 
\sigma$ window around the observed $ \piz $ and $ \eta $ peaks.

Additionally, direct photon candidates have to pass an isolation cut, 
which further reduces the contribution of decay 
photons~\cite{Adare:2012yt}. Ref.~\cite{Adare:2010yw} estimates that the 
contribution of the next-to-leading-order fragmentation photons to the 
isolated direct photon sample is less than 15\% for photons with 
$p_T>5\gevc$. The photon isolation cut requires that the sum of the 
particles' energy surrounding the photon in a cone of radius 
\( r = \sqrt{ (\Delta \eta)^2 + (\Delta \phi)^2 } < 0.4 \) radians be less 
than 10\% of the candidate photon's energy: $E_{\rm cone} < E_\gamma \cdot 10\%$.  
To be included in the cone sum energy, \( E_{\rm cone} \), an EMCal 
cluster must have energy larger than $0.15\gev$ and a charged track needs to 
have a momentum above $0.2\gevc$.  To provide a more inclusive sample of the 
particles surrounding the photon, the clusters and tracks that are included 
in the \( E_{\rm cone} \) sum are only required to pass a minimum set of 
quality cuts. The charged track veto cut is still used to ensure charged 
particles are not double counted by the energy that they deposit in the 
EMCal. The shower-shape cut is not applied to EMCal clusters to ensure that 
neutral hadrons and charged hadrons that were not reconstructed as charged 
tracks can still contribute to \( E_{\rm cone} \).

The asymmetry measurement is formed from photons that satisfy these 
criteria, using similar techniques to previously published PHENIX TSSAs 
which include Refs.~\cite{Aidala:2017pum} and~\cite{upcomingPPG234}.  
The TSSA is determined using the {\it relative luminosity formula}:
\begin{equation}
A_N = \frac{1}{P \, \left<\cos(\phi)\right>} \frac{{N^{\uparrow}}-\mathcal{R}{N^{\downarrow}} }{{N^{\uparrow}}+\mathcal{R}{N^{\downarrow}}},
\label{Equation:RealtiveLuminosity}
\end{equation}

\noindent where \( \mathcal{R} = \mathcal{L}^\uparrow / 
\mathcal{L}^\downarrow \) is the relative luminosity of collisions for 
when the beam was polarized up versus down.  \( P \) is the average 
polarization of the beam and \( \left<\cos(\phi)\right> \) is the 
acceptance factor accounting for the azimuthal coverage of each 
detector arm.  In \eq{Equation:RealtiveLuminosity}, \( N \) refers to 
the particle yield and the up ($\uparrow$) or down ($\downarrow$) arrow 
superscripts refer to the direction of the beam polarization.  The 
asymmetries are calculated separately for each arm of the detector and 
averaged together for the final result, weighted by the statistical 
uncertainty.

The main source of direct-photon background comes from decay photons 
that were not eliminated by the tagging cut because their partner photon 
was not measured.  This can occur because the partner photon was out of 
acceptance, hit a dead area of the detector, or did not pass the 
minimum-energy cut. To calculate the isolated direct-photon asymmetry, 
$A_N^{\rm dir}$, the candidate isolated direct-photon asymmetry, 
$A_N^{\rm iso}$, must be corrected for the contribution from 
background:
\begin{equation} 
    A_N^{\rm dir} = \frac{A_N^{\rm iso}  
                    - r_{\pi^0} \, A_N^{{\rm iso}, \pi^0} 
                    - r_{\eta} \, A_N^{{\rm iso}, \eta} }
                     {1 - r_{\pi^0} - r_{\eta}}.
\label{Equation:BackgroundSubtraction}
\end{equation} 

\noindent This expression removes the effects of background asymmetries 
from isolated \piz photons, $A_N^{{\rm iso}, \pi^0}$, and isolated 
$\eta$ photons, $A_N^{{\rm iso}, \eta}$, where $r_{\pi^0}$ and 
$r_{\eta}$ are the background fractions due to photons from \piz and 
$\eta$ decays, respectively. Because the midrapidity \piz and $\eta$ 
TSSAs have been measured to be consistent with zero to high statistical 
precision~\cite{upcomingPPG234} and their isolated asymmetries were also 
confirmed to be consistent with zero, $A_{N}^{\rm{iso}, \pi^0}$ and 
$A_{N}^{\rm{iso}, \eta}$ are set to zero in 
\eq{Equation:BackgroundSubtraction}.   The systematic uncertainty due to 
setting the background asymmetries to zero dominates the total 
systematic uncertainty of the direct-photon asymmetry for all \pt bins.  
It is assigned by integrating the inclusive midrapidity \piz and $\eta$
TSSAs over photon \pt and propagating their uncertainties through 
\eq{Equation:BackgroundSubtraction}.

\begin{figure}[tbh]
  \centering
  \includegraphics[width=1.0\linewidth]{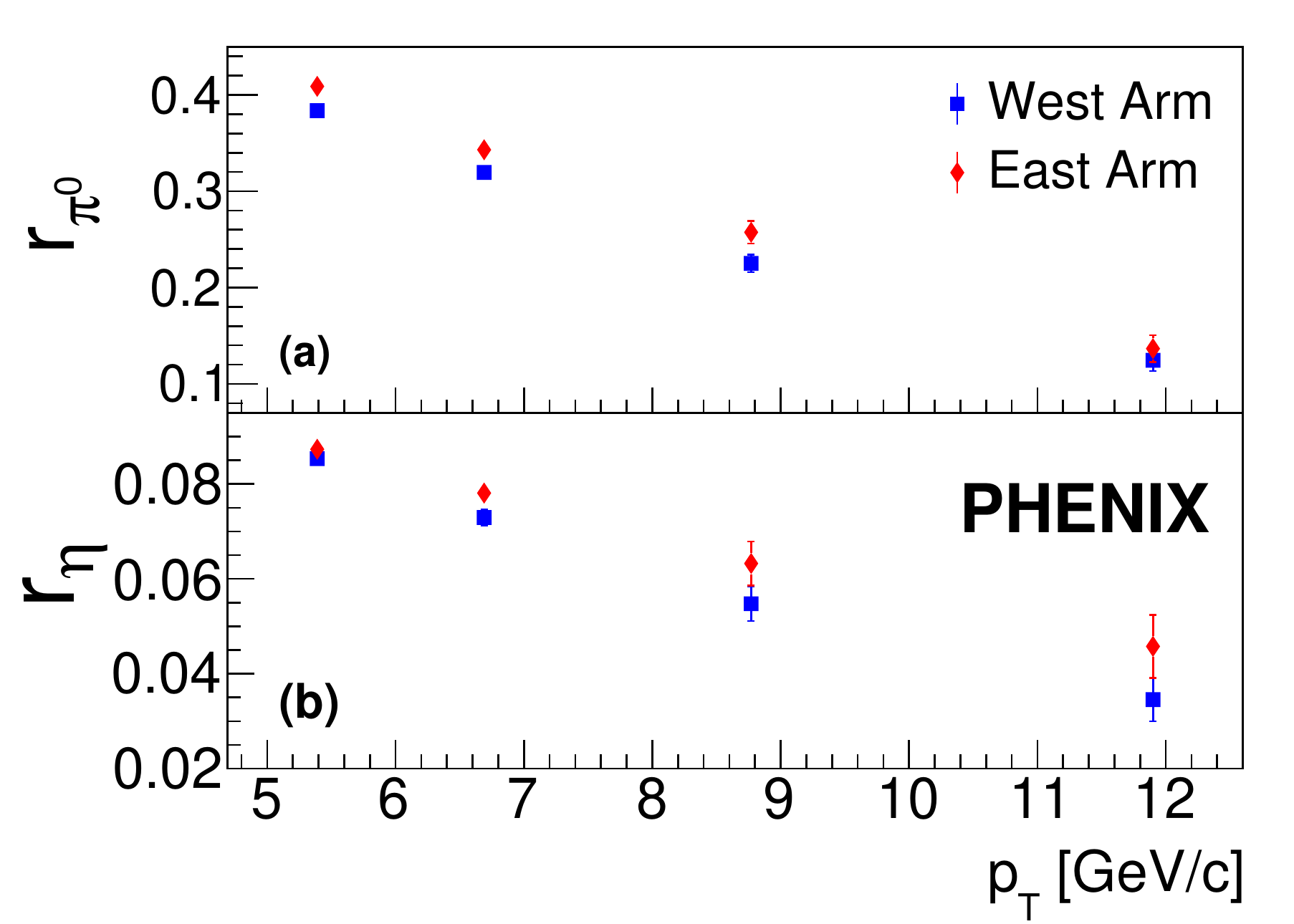}
  \caption{The fractional contribution of photons from (a) \piz and 
(b) $\eta$ decays to the isolated direct photon  candidate sample. }
  \label{fig:r}
\end{figure}

The background fraction calculation is performed by taking the ratio of 
measured photon yields: \( N^{{\rm iso}, h}_{\rm tag} / N^{\rm iso} \), 
where \( N^{\rm iso} \) is the isolated direct photon candidate sample.  
\( N^{{\rm iso}, h}_{\rm tag} \) is the number of photons that were 
tagged as coming from a diphoton decay of hadron \( h \) and pass the 
photon pair isolation cut, $ E_{\rm cone} - E_{\rm partner} < E_\gamma 
\cdot 10\% $, which subtracts off the energy of the partner photon, \( 
E_{\rm partner} \). Tagged photons that pass this cut would have been 
included in the isolated direct photon candidate sample had their 
partner photon not been detected.  Simulations are used to calculate how 
to convert from the number of tagged decay photons to the number of 
decay photons where the partner photon was missed. The background 
fraction, $r_h$, for photons from \piz and \( \eta \) meson decays is 
calculated separately to account for their differences in particle 
production and decay kinematics,
\begin{equation}
r_h = R_h \frac{ N^{{\rm iso}, h}_{\rm tag} }{ N^{\rm iso} },
\label{Equation:BackgroundFraction}
\end{equation} 

\noindent where $R_h$ is the one-miss ratio for the decay of hadron $h$. 
It is the ratio in single particle Monte Carlo of the number of photons 
for which only one of the simulated decay photons was reconstructed to 
the number of photons in which both decay photons were 
reconstructed~\cite{Adare:2012yt}.  These simulations include the 
geometry, resolution, and configuration of the dead areas of the EMCal 
and use the previously measured \piz~\cite{Adare:2007dg} and \( \eta 
\)~\cite{Adare:2010cy} cross sections.  The background fractions for 
photons from \piz(\(\eta \)) decays are plotted in \fig{fig:r} and are 
systematically larger in the east arm versus the west due to the PbGl 
sectors having slightly more dead area compared to the PbSc sectors.  
The contribution of decay photons from sources heavier than
$\eta$ mesons is estimated to be less than 3\% with respect to the 
measured background and so an even smaller percentage of the total direct 
photon sample. The uncertainty on the background fraction is propagated 
through \eq{Equation:BackgroundSubtraction} to assign an additional 
systematic uncertainty to the direct-photon asymmetry.

A similar method to \eq{Equation:BackgroundFraction} is used to find the 
contribution of merged \piz decay photons.  The equivalent $R_h$ is 
calculated using simulated $h \rightarrow \gamma \gamma$ decays, taking 
the ratio of the number of reconstructed EMCal clusters produced by 
merged decay photons divided by the number of reconstructed clusters 
associated with a single decay photon.  The contribution from merged 
photon clusters was found to be less than 0.2\%, small compared to the 
up to 50\% background fraction due to the one-miss effects, and the 
contribution from merged \( \eta \) decays was confirmed to be 
negligible.

An additional systematic study is performed by calculating the asymmetry 
with the {\it square root formula}:
\begin{equation}
A_N = \frac{1}{P \, \left<\cos(\phi)\right>} \frac{\sqrt{N_L^{\uparrow}N_R^{\downarrow}}-\sqrt{N_L^{\downarrow}N_R^{\uparrow}} }{\sqrt{N_L^{\uparrow}N_R^{\downarrow}}+\sqrt{N_L^{\downarrow}N_R^{\uparrow}}},
\label{eq:sqrtan}
\end{equation} 

\noindent where the \( L \) and \( R \) subscripts refer to yields to 
the left and to the right of the polarized-beam-going direction, 
respectively.  This result is verified to be consistent with the 
relative luminosity formula results from 
\eq{Equation:RealtiveLuminosity} and the differences between these 
results are assigned as an additional systematic uncertainty due to 
possible variations in detector performance and beam conditions.  
The systematic uncertainty due to setting the background asymmetries to 
zero dominates the total systematic uncertainty by an order of magnitude 
for all \pt bins except for the highest \pt bin, where it is only slightly
larger than the difference between the {\it square root formula} and {\it 
relative luminosity formula}. Another study using bunch shuffling found no
additional systematic effects.  Bunch shuffling is a technique that 
randomizes the bunch-by-bunch beam polarization directions to confirm that
the variations present in the data are consistent with what is expected by
statistical variation.

\begin{figure}[th]
  \centering
  \includegraphics[width=1.0\linewidth]{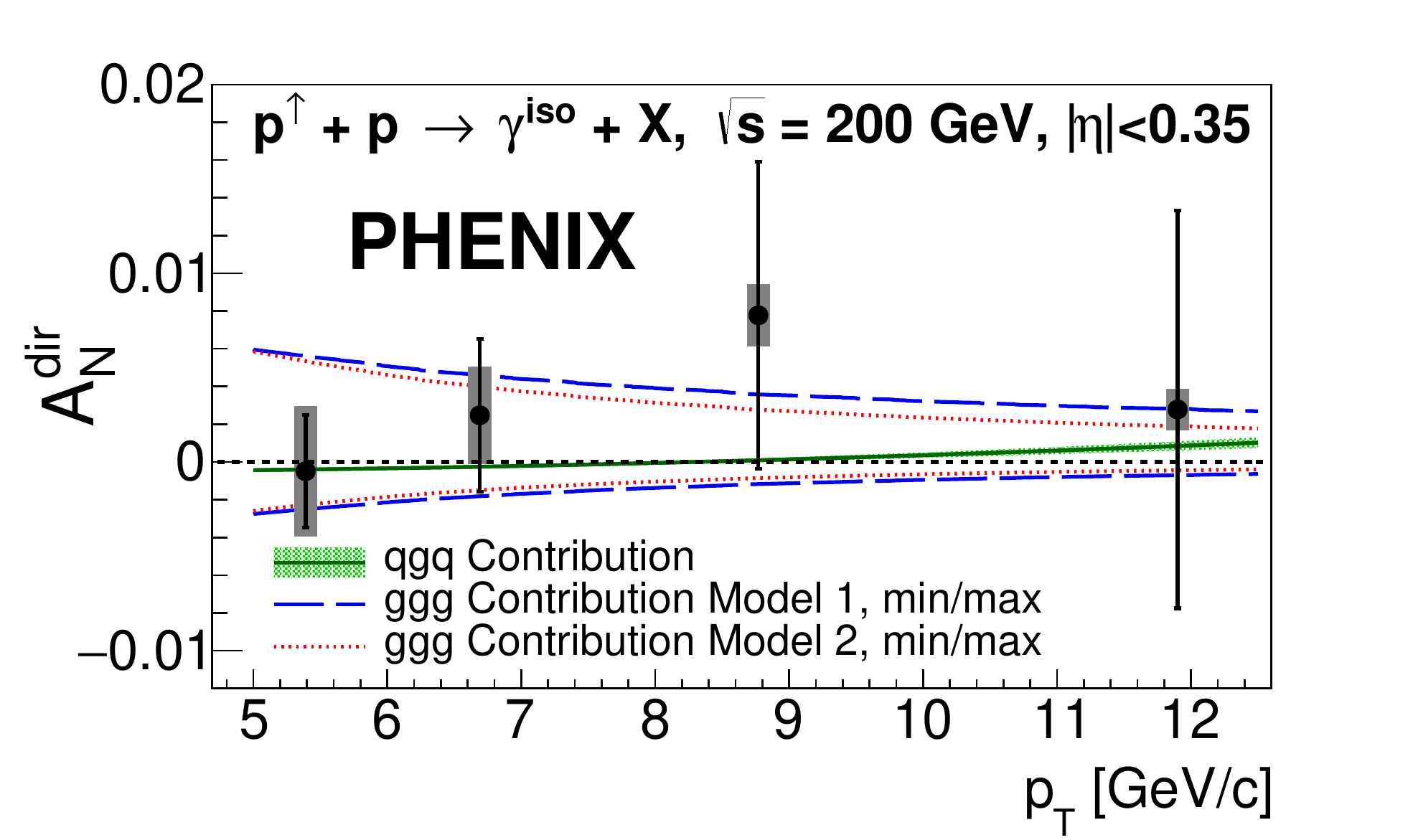}
  \caption{Transverse single-spin asymmetry of isolated direct photons 
measured at midrapidity $|\eta|<0.35$ in \ptp collisions at \sqs = 
200\gev. An additional scale uncertainty of 3.4\% due to the 
polarization uncertainty is not shown. }
  \label{fig:dpANtheory}
\end{figure}

\begin{table}[hb]
 \caption{\label{table:pi0TSSA}
The measured $A_N$ of isolated direct photons in \ptp collisions at 
\sqs=200~GeV as a function of \pt.  An additional scale uncertainty of 
3.4\% due to the polarization uncertainty is not included.
}
\label{tab:an}
 \begin{ruledtabular}
 \begin{tabular}{cccc}
  $\langle\pt\rangle [{\rm GeV}/c]$ & $A_N^{\rm dir}$ & $\sigma_{\rm stat}$ & $\sigma_{\rm syst}$ \\\hline
	 5.39 & -0.000492 & 0.00299 & 0.00341 \\
	 6.69 &  0.00247  & 0.00404 & 0.00252 \\
	 8.77 &  0.00777  & 0.00814 & 0.00159 \\
	11.88 &  0.00278  & 0.0105  & 0.00106 \\
 \end{tabular}
 \end{ruledtabular}
 \end{table}

The results for the $A_{N}$ of isolated direct photons, 
$A_{N}^{\rm dir}$, at midrapidity in \ptp collisions at \sqs = 200\gev\, 
are shown in Table~\ref{table:pi0TSSA} and in \fig{fig:dpANtheory}, 
where the shaded [gray] bands represent the systematic uncertainty and 
the vertical bars represent the statistical uncertainty. The measurement 
is consistent with zero to within 1\% across the entire \pt range. 
\Fig{fig:dpANtheory} also shows predictions from collinear twist-3 
correlation functions.  The solid [green] curve shows the contribution of 
\textit{qgq} correlation functions to the direct-photon asymmetry which 
is calculated using functions that were published in 
Ref.~\cite{Kanazawa:2014nea} that are integrated over the $|\eta| < 0.35$ 
pseudorapidity range of the PHENIX central arms.  This calculation 
includes contributions from the \textit{qgq} correlation functions 
present in both the polarized and unpolarized proton, including the ETQS 
function which is extracted from a global fit in 
Ref.~\cite{Cammarota:2020qcw}.  The error band plotted with the 
solid [green] curve in \fig{fig:dpANtheory} includes uncertainties 
propagated from fits to data, but does not include uncertainties 
associated with assuming functional forms.  Quark-flavor dependence is not
considered in these calculations, including \textit{qgq} correlators.  
Direct-photon production in \pp collisions is four times more sensitive to
the up quark than the down quark in the proton because of the factor of 
electric charge squared in the production cross section.

Given the small predicted contributions from \textit{qgq} correlation 
functions to the midrapidity direct photon TSSA, this measurement can 
provide a clean extraction of the \textit{ggg} function.  The predicted 
ranges for the trigluon correlation function's contribution to the 
direct-photon asymmetry are also plotted in \fig{fig:dpANtheory}. The dashed [blue] and dotted [red] curves use results that were published in 
Ref.~\cite{Koike:2011mb} and were reevaluated as a function of photon \pt 
for pseudorapidity \( \eta = 0 \)~\footnote{The trigluon Model 1 and Model
2 curves in Fig. 2 were provided by S. Yoshida, while D. Pitonyak provided
the quark-gluon-quark curve.}.  Models 1 and 2 assume different functional
forms for the trigluon correlation function in terms of the collinear 
leading-twist gluon PDF; no uncertainties are available for these curves.
As this figure shows, this measurement has the statistical precision, 
especially at low \pt, to constrain the trigluon correlation function.

In summary, the TSSA of midrapidity isolated direct photons was measured 
by the PHENIX experiment to be consistent with zero in the presented \pt 
range, with uncertainties as low as 0.4\% in the lowest \pt bins. This 
is the first time direct photons have been used to probe transversely 
polarized proton collisions at RHIC and the first measurement of this 
TSSA in almost 30 years, with significantly higher \pt reach and up to a 
fifty-fold improvement in uncertainty.  Direct photons are a clean probe 
of proton structure with no contributions from final-state QCD effects 
and at midrapidity are particularly sensitive to gluon dynamics.  When 
included in the global analysis of world TSSA data, this measurement 
will constrain gluon spin-momentum correlations in the transversely 
polarized proton for $x \approx x_T = 0.05 - 0.18$, marking an important
step toward creating a more three-dimensional picture of proton structure.




\begin{acknowledgments}

We thank the staff of the Collider-Accelerator and Physics
Departments at Brookhaven National Laboratory and the staff of
the other PHENIX participating institutions for their vital
contributions.  We also thank D. Pitonyak and S. Yoshida for
helpful discussions.
We acknowledge support from the Office of Nuclear Physics in the
Office of Science of the Department of Energy,
the National Science Foundation,
Abilene Christian University Research Council,
Research Foundation of SUNY, and
Dean of the College of Arts and Sciences, Vanderbilt University
(U.S.A),
Ministry of Education, Culture, Sports, Science, and Technology
and the Japan Society for the Promotion of Science (Japan),
Conselho Nacional de Desenvolvimento Cient\'{\i}fico e
Tecnol{\'o}gico and Funda\c c{\~a}o de Amparo {\`a} Pesquisa do
Estado de S{\~a}o Paulo (Brazil),
Natural Science Foundation of China (People's Republic of China),
Croatian Science Foundation and
Ministry of Science and Education (Croatia),
Ministry of Education, Youth and Sports (Czech Republic),
Centre National de la Recherche Scientifique, Commissariat
{\`a} l'{\'E}nergie Atomique, and Institut National de Physique
Nucl{\'e}aire et de Physique des Particules (France),
Bundesministerium f\"ur Bildung und Forschung, Deutscher Akademischer
Austausch Dienst, and Alexander von Humboldt Stiftung (Germany),
J. Bolyai Research Scholarship, EFOP, the New National Excellence
Program ({\'U}NKP), NKFIH, and OTKA (Hungary),
Department of Atomic Energy and Department of Science and Technology
(India),
Israel Science Foundation (Israel),
Basic Science Research and SRC(CENuM) Programs through NRF
funded by the Ministry of Education and the Ministry of
Science and ICT (Korea).
Physics Department, Lahore University of Management Sciences (Pakistan),
Ministry of Education and Science, Russian Academy of Sciences,
Federal Agency of Atomic Energy (Russia),
VR and Wallenberg Foundation (Sweden),
the U.S. Civilian Research and Development Foundation for the
Independent States of the Former Soviet Union,
the Hungarian American Enterprise Scholarship Fund,
the US-Hungarian Fulbright Foundation,
and the US-Israel Binational Science Foundation.

\end{acknowledgments}



%
 
\end{document}